\documentclass[compsoc,conference,a4paper,10pt,times]{IEEEtran}
\IEEEoverridecommandlockouts
\usepackage{cite}
\usepackage{amsmath,amssymb,amsfonts}
\usepackage{algorithmic}
\usepackage{graphicx}
\usepackage{textcomp}
\usepackage{bmpsize}
\usepackage{xcolor}
\usepackage{lipsum}
\usepackage[colorlinks=true,urlcolor=black]{hyperref}
\def\BibTeX{{\rm B\kern-.05em{\sc i\kern-.025em b}\kern-.08em
    T\kern-.1667em\lower.7ex\hbox{E}\kern-.125emX}}

\usepackage{tikz}
\usepackage{amsmath}

\usepackage{filecontents}

\usepackage{tikz}
\usepackage{amsmath}

\usepackage{wasysym}

\usepackage{algorithmic}
\usepackage{graphicx}
\usepackage{tabularx}
\newcolumntype{Y}{>{\centering\arraybackslash}X}
\usepackage{textcomp}
\usepackage{xcolor}

\usepackage{amsmath, bm}

\usepackage{tikz}
\usetikzlibrary{positioning, shapes.geometric}

\usepackage{amsmath}
\usepackage{tikz}
\usetikzlibrary{shapes, arrows.meta, positioning, calc}

\usepackage{tikz}
\usepackage{pgfplots}

\usepackage{textcomp}
\usepackage{stfloats}
\usepackage{url}
\usepackage{tcolorbox}
\usepackage{verbatim}
\usepackage{graphicx}

\usepackage{amsmath,amssymb,amsfonts}
\usepackage{algorithmic}
\usepackage{graphicx}
\usepackage{textcomp}
\usepackage{xcolor}

\usepackage{amsmath, bm}

\usepackage{soul}

\usepackage{mathtools}
\usepackage{relsize}

\usepackage{booktabs}

\usepackage[para,online,flushleft]{threeparttable}

\usepackage{amsmath}

\usepackage{subcaption}
\usepackage{graphicx}
\usepackage{mathrsfs}  
\usepackage{pifont}

\usepackage{xcolor}
\usepackage[linesnumbered,ruled,vlined]{algorithm2e}

\SetCommentSty{mycommfont}

\SetKwInput{KwInput}{Input}                
\SetKwInput{KwOutput}{Output}              

\usepackage{algorithmic}
\usepackage{graphicx}
\usepackage{amsmath}
\usepackage{textcomp}
\usepackage{xcolor}
\def\BibTeX{{\rm B\kern-.05em{\sc i\kern-.025em b}\kern-.08em
    T\kern-.1667em\lower.7ex\hbox{E}\kern-.125emX}}

\usepackage{adjustbox}
\usepackage{multirow}
\usepackage{amssymb,mathtools}

\usepackage{array}
\newcolumntype{P}[1]{>{\centering\arraybackslash}p{#1}}
\newcolumntype{M}[1]{>{\centering\arraybackslash}m{#1}}

\usepackage{tikz}

\usepackage{xcolor}
\usepackage[linesnumbered,ruled,vlined]{algorithm2e}

\usepackage[symbol]{footmisc}

\usepackage{hyperref}
\usepackage{xcolor}
\usepackage{threeparttable}
\usepackage{booktabs}

\usepackage{eso-pic}

\newtcolorbox{boxA}{
    fontupper = \bf,
    boxrule = 1.5pt,
    colframe = black 
}

\usepackage{soul}    
\usepackage{xcolor}  

\makeatletter
\newcommand{\linebreakand}{%
  \end{@IEEEauthorhalign}
  \hfill\mbox{}\par
  \mbox{}\hfill\begin{@IEEEauthorhalign}
}
\makeatother

\usepackage{listings}

\usepackage[nolist]{acronym}

\usepackage{pdflscape}
\usepackage{afterpage}
\usepackage{capt-of} 

\usepackage[symbol]{footmisc}

\usepackage{hyperref} 

\begin{document}

\title{Automated Stealthy Wear-Out Attack on Digital Twins With Deep Reinforcement Learning}

\author{
\IEEEauthorblockN{Joshua Haworth\textsuperscript{$\boxtimes$}, Aryan Pasikhani, George Pavlides, Prosanta Gope, John Clark}
\IEEEauthorblockA{Department of Computer Science, University of Sheffield, United Kingdom \\
\{jhaworth1, aryan.pasikhani, p.gope, john.clark\}@sheffield.ac.uk, p-george18@hotmail.com}
}

\AddToShipoutPictureFG*{%
  \AtTextUpperLeft{%
    \put(0,20){%
      \makebox[\textwidth][c]{\normalsize This paper has been accepted to appear in the IEEE European Symposium on Security and Privacy (Euro S\&P), 2026.}%
    }%
  }%
}
\maketitle
\begin{acronym}
\acro{DT}{Digital Twin}
\acro{PHM}{Prognostics and Health Management}
\acro{RL}{Reinforcement Learning}
\acro{DRL}{Deep Reinforcement Learning}
\acro{DoS}{Denail of Service}
\acro{AI}{Artificial Intelligence}
\acro{AMRC}{Advanced Manufacturing Research Centre}
\acro{IDS}{Intrusion Detection System}
\acro{MitM}{Man-in-the-Middle}
\acro{PE}{Physical Entity}
\acro{VE}{Virtual Entity}
\acro{EV}{Electric Vehicle}
\acro{SoC}{State of Charge}
\acro{ICS}{Industrial Control System}
\acro{V2G-CPS}{Vehicle-to-Grid Cyber-Physical System}
\acro{FDIA}{False Data Injection Attack}
\acro{SA}{Switch Attack}
\acro{LFC}{Load Frequency Control}
\acro{DDPG}{Deep Deterministic Policy Gradient}
\acro{PPO}{Proximal Policy Optimization}
\acro{AGC}{Automatic Generation Control}
\acro{PMU}{Phasor Measurement Unit}
\acro{S}{Service}
\acro{DD}{DT Data}
\acro{CN}{Connection}
\acro{SAC}{Soft Actor-Critic}
\acro{TD3}{Twin-Delayed Deep Deterministic Policy Gradient}
\acro{A2C}{Advantage Actor-Critic}
\acro{TP}{True Positive}
\acro{TN}{True Negative}
\acro{FP}{False Positive}
\acro{FN}{False Negative}
\acro{LSTM}{Long Short-Term Memory}
\acro{CNN}{Convolutional Neural Network}
\acro{ResNet}{Residual Neural Network}
\acro{GRU}{Gated Recurrent Unit}
\acro{ReLU}{Rectified Linear Unit}
\acro{CPS}{Cyber-Physical System}

\end{acronym} 
\begin{abstract}
Digital Twins (DTs) have emerged as pivotal enablers of Industry 4.0, offering transformative capabilities such as real-time monitoring, advanced simulation, and precise control of physical assets. By bridging the physical and virtual domains, DTs facilitate seamless integration of data-driven decision-making and operational optimisation. However, this seamless interaction significantly expands the attack surface of industrial systems, creating vulnerabilities that adversaries can exploit.
This paper introduces a novel and stealthy wear-out attack leveraging Deep Reinforcement Learning (DRL) to target DT-enabled infrastructures. The adversary strategically and covertly manipulates control signals, inducing increased torque on a specific joint to accelerate wear and tear while evading detection by a state-of-the-art anomaly detection system. Extensive benchmarking of reinforcement learning algorithms - including Twin Delayed Deep Deterministic Policy Gradient (TD3), Soft Actor-Critic (SAC), Proximal Policy Optimisation (PPO), and Advantage Actor-Critic (A2C) - revealed that SAC consistently outperformed its counterparts in terms of sample efficiency, stability, and overall attack effectiveness.
We evaluate the proposed adversary in an industrial setting using the UR10e robotic arm. Results demonstrate the adversary’s ability to significantly elevate torque levels on the targeted joint, leading to accelerated degradation and increased maintenance costs, all while operating stealthily and avoiding detection. Our findings highlight the substantial risks posed by DRL-driven adversaries to DT-enabled environments and emphasise the critical need for robust defence mechanisms to protect critical industrial systems.
\end{abstract}


\vspace{-1em}

\section{Introduction}

Striving to realise the Industry 4.0 vision represents one of the top priorities for manufacturers\footnote{\url{https://www.sap.com/products/scm/industry-4-0/industry-4-0-strategy.html}}. The use of industrial robots, cloud computing and vast amounts of sensor data collected throughout a product's lifecycle are expected to improve production efficiency, flexibility and decision making, while at the same time reducing waste and minimising carbon emissions \cite{i4.0, dt_ind_sota}.
\ac{DT} technology is one of the vital enablers for the realisation of Industry 4.0 \cite{dt_ind_sota}. It provides a virtual replica of a physical entity, system or process, that gets updated in real-time according to past data, real-time sensor readings and a predefined physical model \cite{dt_ind_sota}. \acp{DT} seamlessly integrate and analyse physical and digital data recorded throughout a product's lifecycle, providing new services which can be utilised to adjust how an operation is performed in the physical space according to direct orders from the virtual space \cite{dt_ind_sota}. This can improve the industrial process' performance, enhance product designs and streamline operations, but also allow for \ac{PHM} to detect, in a timely manner, potential asset faults and degradations, thus reducing maintenance costs\cite{dt_ind_sota}.

Tao et al. \cite{TAO2018} proposed a five-dimensional model of \ac{DT} in which there is a clear separation between the elements involved. Specifically, the \ac{DT} architecture is decomposed into \ac{PE}, \ac{VE}, \acp{S}, \ac{DD} and \ac{CN}. As a faithful replica of the \ac{PE}, the \ac{VE}  can be utilised to run simulations or tests of various scenarios without affecting the real-world counterpart. This is particularly beneficial for testing configuration changes before deploying them to critical physical systems or for performing security assessments on systems whose potential downtime is unacceptable \cite{elegant}. \acp{DT} are usually hosted on the cloud, providing them with vast amounts of computational power, which allows them to perform complex calculations and analyses that would otherwise not be feasible on the, usually more resource-constraint, physical counterparts \cite{prototypingIntrusionDetection}. Whilst the benefits of \ac{DT} are clear \cite{dt_ind_sota, elegant}, the two-way communication between the physical and cyber spaces, combined with the introduction of "smart" services, broadens the attack surface and presents new security challenges to the system \cite{prototypingIntrusionDetection}. For example, an attacker that has infiltrated the system or compromised one or more elements of the \ac{DT} model (e.g. \ac{VE}, \acp{S}, \ac{DD} or \ac{CN}) could manipulate the \ac{VE}'s instructions to command the \ac{PE} to perform hostile or unsafe actions. Such actions have the potential of harming assets in the physical space, producing defective products and resulting in significant financial losses for the manufacturer. In case the affected industrial system is part of critical infrastructure, such as a power grid, the consequences of malicious compromise can be of significant scale and even lead to the loss of human life \cite{scadaVuln}.  Moreover, since the \ac{DT} model includes an accurate replica of a physical object and its operational logic, unauthorised access to the \ac{DT}'s elements can lead to the leak of confidential information regarding how processes are performed and sensitive Intellectual Property \cite{fernandez2024surveyprivacyattacksDT}.

With the security risks that arise due to the use of \ac{DT} technology, it is apparent that its operation should be safeguarded via thorough security assessments and the design of appropriate defence measures. A considerable amount of work has been conducted on implementing defence mechanisms on the \ac{DT}-side in order to detect and mitigate attacks that aim to exploit the existence of a \ac{DT} to deceive the physical object into falling into an unsafe state (e.g. \cite{dtCPMSBalta, secFrameworkDTCloud, prototypingIntrusionDetection, smartDTV2G}). However, in the existing literature, the adversary is non-intelligent, and their malignant modifications are constant, monotonous or probabilistic \cite{secFrameworkDTCloud, prototypingIntrusionDetection, akbarian2020, balta2024}, with no concern for intelligently adapting and adjusting their attack strategies based on the observed environment. 

In recent years, \ac{AI}-assisted attacks targeting industrial systems emerged  (e.g. \cite{chungSmartMalware, chen2022}) that are not only highly effective in achieving their hostile objectives but also have high evasion rates against existing defence measures. 
In particular, \ac{DRL} algorithms have demonstrated promising performance as an offensive tool against industrial environments \cite{Mohamed_2024, shereen2024, maiti2023}, capable of synthesising highly potent and stealthy attack strategies. Their ability to learn how to conduct effective attacks with little knowledge of the victim system (a.k.a grey box) or potential attack approaches makes them a versatile tool for discovering new attack strategies \cite{Mohamed_2024}. Furthermore, as a data-driven method that depends on (partial) system observations, \ac{DRL} agents are capable of discovering and exploiting complex state dynamics and inter-dependencies, making them a powerful technique for attacking convoluted systems \cite{Mohamed_2024, maiti2023}. Also, the fact that \ac{DRL} agents take strategic actions that aim to maximise the \textbf{cumulative reward} received enables them to perform better in the long run compared to alternative short-sighted attack methods \cite{li2022learning}. 

So far, no published research investigates the possibility of utilising \ac{AI}-based methods to exploit and attack \ac{DT}-enabled infrastructure. With the increasing adoption of \ac{DT} technology by world leaders in various fields like power grids, wastewater management, car manufacturing,
aerospace engineering, and healthcare \cite{dt_ind_sota}, safeguarding \ac{DT}-enabled infrastructure is more relevant than ever before. Any compromise in their intended functionality has the potential of a wide negative impact on manufacturers and customers across the world.  With the rising occurrence and sophistication of damaging attacks against industrial environments (e.g. the Stuxnet attack against Iran's nuclear facilities\cite{stuxnet}, the BlackEnergy malware attack on Ukraine's power grid \footnote{\url{https://www.cisa.gov/news-events/ics-alerts/ir-alert-h-16-056-01}} and the HatMan malware targeting Schneider Electric safety controllers \footnote{\url{https://www.cisa.gov/sites/default/files/documents/MAR-17-352-01\%20HatMan\%E2\%80\%94Safety\%20System\%20Targeted\%20Malware_S508C.pdf}}), it is apparent that industrial systems are a sought-after target by cyber criminals and need to be secured. By proactively identifying attack vectors, we can inform and facilitate the development of effective and robust defence measures \cite{Mohamed_2024}.

\subsection{Our Contribution}

In this paper, we employ the power of \ac{DRL} to train an adversarial agent capable of conducting stealthy wear-out attacks against \ac{DT}-enabled infrastructure. In our specific use case, the \ac{DRL} agent attempts to make subtle changes to the control signals issued by the \ac{VE} for a physical robotic arm in order to gradually wear out a selected joint of it. This leads to the physical system's faster deterioration, degrades operational efficiency and effectiveness, and increases the fault occurrence probability and the owner's maintenance and replacement costs.

In summary, this paper's contributions are:
\begin{itemize}
    \item We develop the \textit{first} \ac{DRL}-based adversary who can model the robotic arm's safe zones of operation via the \ac{DT}, and perform an intelligent wear-out attack on the targeted joint. Here, we consider a grey-box attacker by altering the access to environment observations the agent has. 

\item We demonstrate how our novel DRL-assisted adversary can autonomously manipulate robotic operations and induce wear-out effects by employing a stealthy low-and-slow attack strategy. The adversary effectively evades detection by an ensemble of Autoencoder-based anomaly detectors, highlighting its capability to exploit vulnerabilities over prolonged periods.

\item We publicly share our implementation code along with the generated adversarial policies and results\footnote{\textbf{All related materials, including datasets and code, are available for the research community here: \url{https://anonymous.4open.science/r/Stealthy-Wear-Out-RL-BC8B/README.md}}} for reproducibility purposes and also to facilitate their use by the research community for the development of robust and resilient defence systems.
\end{itemize}

To the best of our knowledge, no work currently exists that investigates the possibility of a stealthy wear-out attack against physical assets in a \ac{DT}-enabled setting via the use of \ac{DRL}.

\begin{figure*}[htbp] 
\center{\includegraphics[width=165mm]{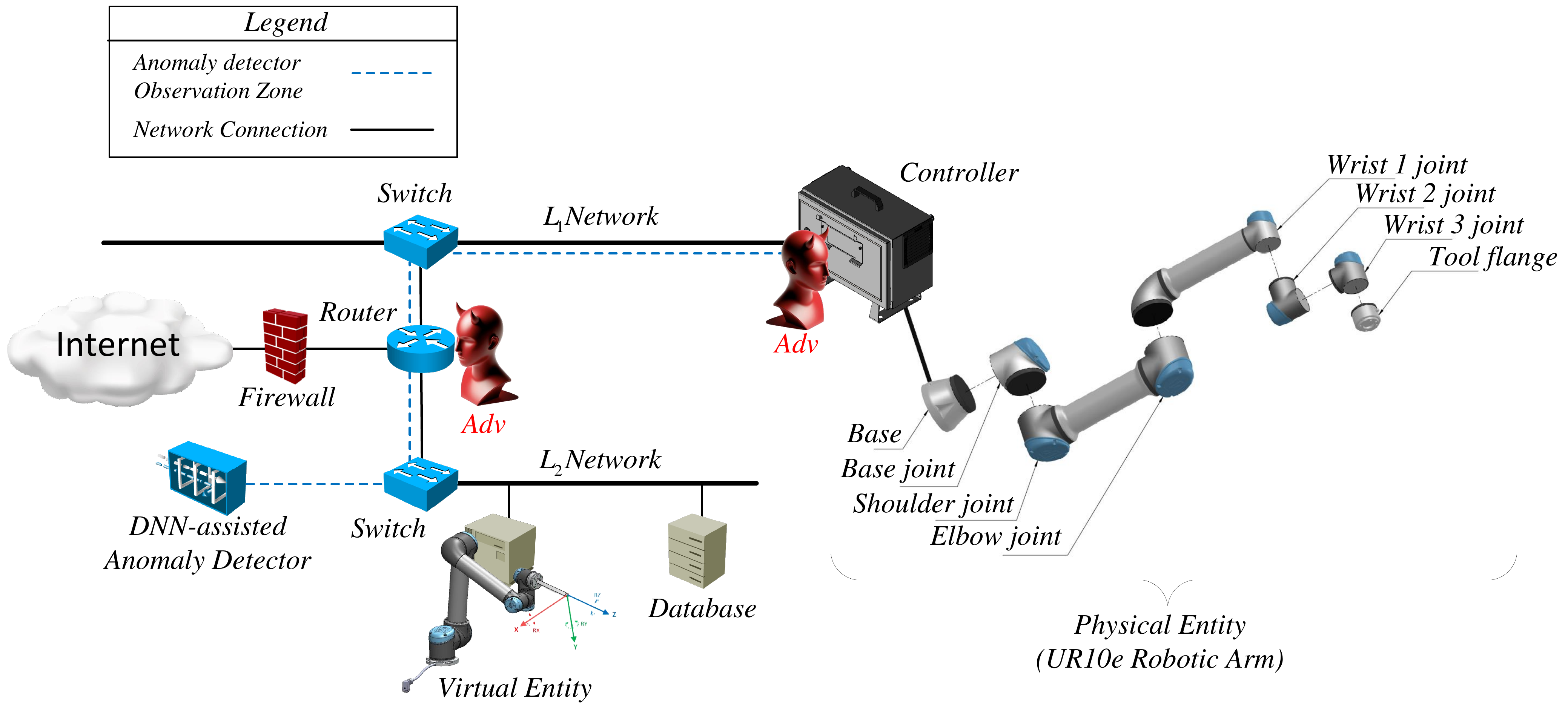}} 

\caption{DRL-assisted adversary targeting an industrial robotic system by exploiting vulnerabilities in $L_1$ (Operational Technology) or $L_2$ (Information Technology) networks. }
\label{fig:Attack Surface} 
\end{figure*}

\section{Related Work}
\label{sec:RelatedWork}

Digital Twins (DTs) are becoming integral to modern \ac{ICS}, enabling tighter coupling between physical processes and digital operations. As industrial environments adopt greater automation and connectivity, \acp{DT} provide continuous synchronisation between virtual and physical entities, supporting real-time monitoring, prediction, and control \cite{dt_ind_sota, TAO2018}. Their growing deployment across manufacturing, energy, and transportation has shifted \acp{DT} from auxiliary analytical tools to essential components of Industry 4.0 infrastructures \cite{i4.0,dt_ind_sota}.

\textbf{Attacks against \ac{DT}-enabled infrastructure:} 
As \acp{DT} become increasingly adopted within industrial settings, consideration needs to be given to new attack vectors introduced through bidirectional synchronisation. The existing literature focuses on adversaries that introduce predefined manipulations to the communication between the \ac{PE} and \ac{VE}. Early work by Eckhart and Ekelhart \cite{Eckhart2018} explore state synchronisation vulnerabilities in \ac{CPS}, modelling an insider and \ac{MitM} adversaries capable of spoofing or modifying direct communication between the \ac{VE}, and \ac{PE}. The attack presented relies on the direct manipulation of values without incorporating stealth or situational adaptivity, demonstrating an initial focus on basic intrusion scenarios within DT-enabled environments. Subsequent works by Akbarian et al \cite{akbarian2020}, and Tärneberg et al \cite{prototypingIntrusionDetection} explore more advanced manipulation techniques with the additional focus of attempting to bypass \ac{IDS} mechanisms within the system. In \cite{akbarian2020} ramping and scaling attacks are introduced as an alternative to constant signal offset, such that scaling attacks rely on a scaling factor used to manipulate the original measurement value, and ramping attacks introduce a gradually increasing manipulation. Tärneberg et al \cite{prototypingIntrusionDetection} build on these attack methods with the introduction of burst attacks where the adversarial manipulations are applied in ON/OFF periods that determine when the attack occurs, with the aim to reduce detectability. Both of these works demonstrate how minor manipulations can effect the system behaviour while attempting to avoid security mechanisms, highlighting the potential consequences of data injection attacks against \ac{DT} systems.

Dietz et al \cite{dietz2022employing} further formalise adversarial capabilities in a DT-enabled industrial systems, simulated via virtual machines, using the Dolev-Yao attacker model \cite{dolev-yao}. This assumes secure cryptographic primitives, full system knowledge, and possesses the ability to run infinite concurrent processes. Varghese et al \cite{varghese2022digital} use the same basis outlined in \cite{dietz2022employing}, with consideration for a range of attacks, including network \ac{DoS}, command injection, and calculated (small factor scaling) and naive measurement modifications (constant / random modifications). More recently Ali et al \cite{Ali_etal_2023} investigate \acp{DT} in an \ac{EV} environment, where the attacker is limited to the manipulation of the voltage and phase angle. By injecting adjustments to the in-transit information vector the adversary aims to corrupt the system state while limiting the adjustments to avoid detection.

A notable limitation of these \acp{FDIA} \cite{secFrameworkDTCloud, prototypingIntrusionDetection, akbarian2020, Ali_etal_2023, dietz2022employing, varghese2022digital, Eckhart2018} is there reliance on manually tuned parameters. In order for these attacks to achieve any semblance of success, the attacker must explicitly calibrate the parameters used to make manipulations for each system in order to balance impact and stealth. In contrast, an intelligent \ac{AI}-based adversary could dynamically adapt its strategy in order to optimise impact and undetectability in response to the environment. This adaptability makes adversarial agents a prime candidate for overcoming the weaknesses of existing attacks against \ac{DT}-enabled infrastructure. Additionally it should be noted that current work focuses on the deployment of security mechanisms as opposed to the use of advanced adversarial models to improve existing security mechanisms.

\textbf{(D)RL-based attacks against industrial systems:} Beyond DT-enabled environments, \ac{DRL} agents are being used to perform attacks against industrial and power grid systems. Mohamed and Kundur \cite{Mohamed_2024} employ \ac{DRL} to synthesise attacks against load frequency control used by power grids. The adversarial agent aims to trigger unwanted protection mechanisms which can lead to sudden power imbalance, grid instability and subsequently, blackout. They demonstrate that an \ac{DRL} agent is capable of generating attacks with little to zero prior knowledge of the victim system and is even able to adjust its attacks to damage different systems. Their agent executes \ac{FDIA} (corrupting measurements and control signals) as well as load switch attacks to destabilise the target grid. Using the \ac{DDPG} \ac{DRL} algorithm, due to its simplicity and support for continuous action and observation spaces, they were able to achieve near-optimal results in respect to the effectiveness of attacks, with the authors claiming that additional training episodes could lead to optimality.

A number of studies demonstrated the use of (D)RL to synthesise attack strategies against power grid systems. In \cite{yan2017, ni2017, wang2021}, the authors use (deep) Q-Learning to perform line-switching attacks that induce sudden changes in grid topology and cause cascading failures, leading to blackouts. Chen et al. \cite{chen2019} trained a Q-Learning With Nearest Sequence Memory agent that, with little knowledge of the victim system, is capable of performing stealthy \acp{FDIA}, modifying system measurements to deceive existing controls in triggering power outages. In \cite{abianeh2022}, a multi-agent \ac{DRL} algorithm using \ac{DDPG} is proposed that is capable of generating stealthy, coordinated \acp{FDIA} against microgrids, which were shown to be effective against a state-of-the-art detection scheme. Shereen et al.  \cite{shereen2024} proposed the use of \ac{PPO} \ac{DRL} algorithm to discover efficient but subtle attack policies that inject false measurements in the readings transmitted to automatic generation control, causing it to issue inaccurate control commands to the generators that can have catastrophic consequences for the power system. In a more complex work, \cite{maiti2023} introduce a Monte Carlo Simulation to identify the most vulnerable operation intervals of a power grid and use a \ac{DDPG} agent to launch scaling attacks on the power flow sensor readings sent to the controlling automatic generation controller during those vulnerable time periods, thus destabilising the grid. To obstruct the attack's detection they also launch coordinated GPS timestamp spoofing attacks on the phaser measurement unit data.

As shown, the use of \ac{DT} in environments can lead to severe cyberattacks, and (D)RL can be a potent tool for conducting attacks against industrial systems. However, to the best of our knowledge, \ac{DRL} has not yet been applied to maliciously exploit the presence of \ac{DT} in an infrastructure, and particularly to perform stealthy wear-out attacks against the infrastructure's physical systems.

\section{Threat Modelling}\label{sec:ThreatModelling}

\noindent\textbf{Adversary's Objectives.}
The adversary aims to covertly manipulate the control signals transmitted from the \ac{VE} to the \ac{PE} in order to achieve long-term physical degradation of the targeted joint(s) of the robotic system. This can be defined by \textbf{three main objectives}:

\begin{itemize}
    \item \textbf{Accelerate Mechanical Wear-Out:} Gradual physical degradation achieved by subtly manipulating the waypoint coordinates transmitted to the robotic arm, thereby inducing higher joint torque and mechanical stress. Sustained increases in load accelerate wear, reducing component lifespan and ultimately raising maintenance costs.
    
    \item \textbf{Undetectability within the Network:} The adversary aims to remain undetected by both the \ac{DT}’s anomaly detection mechanisms (e.g., autoencoder-based detectors) and any built-in safety monitors. By applying small perturbations to waypoint commands, the attacker avoids triggering alarms while still imparting cumulative mechanical stress.

    \item \textbf{Persistence in the System:} Instead of inducing an immediate malfunction, the adversary employs repeated minute modifications that gradually erode system reliability. This “low and slow” approach requires continual adaptation to operational conditions, enabling maximised long-term wear-out while remaining inconspicuous.
\end{itemize}

\noindent\textbf{Adversary's Capabilities.}

Figure \ref{fig:Attack Surface} depicts the attack surface targeted by our DRL-assisted adversary within a DT-enabled control system. The adversary resides on the communication path between the \ac{VE} and the \ac{PE}, enabling full interception, manipulation, and injection of control traffic consistent with the Dolev–Yao threat model \cite{dolev-yao}. From this vantage point, the adversary can alter waypoint commands (control signals) as well as telemetry information exchanged between components, including joint positions, velocities, and accelerations.


\begin{tcolorbox}
    \textbf{Key Assumption}: We assume the adversary can access either $L_1$ or $L_2$, but not both. Simultaneous access enables direct manipulation of the PE while evading the \ac{IDS} through controller-level adjustments. Notably, if the adversary gains $L_1$ access, modifications can be applied directly rather than through control signals sent from the \ac{VE}.
\end{tcolorbox}

Our proposed attack aligns with the \textit{Exploitation} and \textit{Actions on Objectives} phases of the Cyber Kill Chain \footnote[2]{\url{https://www.lockheedmartin.com/en-us/capabilities/cyber/cyber-kill-chain.html}}, where the adversary gathers intelligence on system vulnerabilities and executes targeted disruptions to achieve its malicious objectives while maintaining operational stealth.

We assume the adversary has compromised the VE–PE communication pipeline - for example, via malware leveraging publicly disclosed vulnerabilities such as \textbf{CVE-2024-2442} (exposing ICS protocol interfaces), \textbf{CVE-2024-2882} (enabling unauthorised code execution and data manipulation), or \textbf{CVE-2023-5885} (facilitating persistent command-and-control). Such compromise may occur via human–machine interfaces at Level 1 or through a remote gateway associated with \ac{DT} services. Additionally, we assume that the communication between the \ac{PE} and \ac{VE} occurs using widely used industrial systems protocols such as MODBUS-TCP, which have known and documented vulnerabilities (for example \textbf{CVE-2025-62578}- data is transferred in plaintext and \textbf{CVE-2025-48466} - which enables attackers to send command packets remotely). Following infiltration, the adversary deploys a DRL-based policy that selectively manipulates waypoint commands transmitted from the \ac{VE} to the \ac{PE}.

The adversary introduces small, per-axis modifications to reshape the trajectory without causing overtly invalid or unreachable poses. These modifications are formally defined through:

\begin{equation}
    [\Delta x, \Delta y, \Delta z] = [x, y, z] + [m_x, m_y, m_z]
\end{equation}

Where $[x, y, z]$ represents the original waypoint, $[m_x, m_y, m_z]$ forms the modification made in each axis, and $[\Delta x, \Delta y, \Delta z]$ is the manipulated waypoint received by the \ac{PE}. The modification range was determined based on preliminary experiments, which showed that UR10e can reach waypoints with coordinates (x, y, z) in the range [-1, 1]. Since one of the main requirements of this study is that the agent's actions be stealthy, the modification's range was set to 10\% of the arm's reach. This way, it provides the agent with an action space that is at once flexible enough to explore damaging strategies and constrained enough to facilitate convergence to a stealthy policy.

Our work assumes a \emph{grey-box} threat model, in which the adversary has partial visibility into the DT system, specifically, limited knowledge of the feature set and operational telemetry reflected in the observation space. Such information may stem from system logs, publicly available documentation (e.g., technical specifications), or prior reconnaissance. With this partial insight, the adversary can craft subtle, targeted perturbations that remain within expected operational bounds, improving their ability to evade anomaly detection while exerting strategic influence over the system’s control behaviour.

\textbf{Adversary's Observation}
In the context of a Universal Robots UR10e \footnote[3]{\url{https://www.universal-robots.com/products/ur10e/}} system, the adversary has access to six joints’ runtime data, reflecting the arm’s standard degrees of freedom. Specifically: 
\begin{itemize} 
\item \textbf{\emph{Original Waypoints:}} The baseline coordinates generated by the \ac{DT} (or human operator) for each motion segment. 
\item \textbf{\emph{Modified Waypoints:}} The altered waypoint coordinates as they were executed at the previous step (or “episode”) after the addition of the adversarial offsets.  
\item \textbf{\emph{Joint States:}} Position, velocity, and acceleration for all six joints as measured at the previous step (or “episode”). These provide feedback on how past perturbations affected actual system dynamics and how close the manipulations were to triggering alarms (via the reward function). 
\end{itemize} 

Fig.~\ref{fig:Attack Surface} schematically illustrates how the adversary sits “in the loop”, monitoring and adjusting these data streams. In an out-of-context scenario (e.g., UR5 or a different brand of robot with a different number of joints), the attacker must adapt its policy to whichever number of degrees of freedom the new arm supports but can follow the same methodology: intercepting digital commands, applying minimal offsets, and observing sensor readouts to refine the DRL policy.

\smallskip Overall, through minimal real-time modifications of waypoint commands—powered by DRL training and guided by stealth feedback—the adversary accomplishes sustained, covert wear-out of a targeted joint in the UR10e system or any similarly configured DT-driven robotic platform.

\subsection{Formalisation of Attack} \label{sec:formalisation-pomdp-td3}

The adversarial strategy operates in a Partially Observable Markov Decision Process (POMDP),  under the assumption that the adversary may not have complete observability of the environment. For instance, certain joint readings (e.g., velocity and acceleration) may be inaccessible, reflecting real-world scenarios where the adversary has limited visibility into the system state. The POMDP is defined as the tuple $(S, A, T, R, X, O, \gamma)$, where:
\begin{itemize}
    \item $S$: The set of all possible system states, with $s_t \in S$ representing the system's underlying state at time $t$.
    \item $A$: The set of actions available to the adversary, where $a_t \in A$ represents perturbations to waypoint coordinates. The per-axis modifying actions follow $\Delta_{x}, \Delta_{y}, \Delta_{z} \in [-0.1,0.1] $. 
    \item $T$: The transition probability function $T(s_{t+1} | s_t, a_t)$, describing the likelihood of transitioning to state $s_{t+1}$ from $s_t$ after taking action $a_t$.
    \item $R$: The reward function $R(s_t, a_t, s_{t+1})$, quantifying the adversary’s success in causing increased mechanical stress while avoiding detection.
    \item $X$: The observation space, where $x_t \in X$ represents the adversary’s partial view of the system.
    \item $O$: The observation probability function $O(x_t | s_t)$, describing the likelihood of observing $x_t$ given the true state $s_t$.
    \item $\gamma$: The discount factor ($0 \leq \gamma < 1$), prioritizing immediate rewards over future ones.
\end{itemize}

The adversary’s policy $\pi_\phi: X \rightarrow A$ maps observations to actions and is optimized to maximize the expected cumulative reward:
\begin{equation}
\pi^* = \arg\max_{\pi} \mathbb{E}_{\pi} \left[ \sum_{t=0}^\infty \gamma^t R(s_t, a_t, s_{t+1}) \right].
\end{equation}

\textbf{Algorithm:}
To address partial observability, we designed our adversary based on \cite{SAC} to operate on observations $x_t$ rather than full states $s_t$. The algorithm incorporates the following modifications:

\begin{enumerate}
    \item \textbf{Observation-Based Policy:} The adversary selects actions using a deterministic policy with inherent Gaussian noise for policy exploration:
    \begin{equation}
        a_t = f_\phi (\varepsilon_t;x_t), \quad \varepsilon \sim \mathcal{N}(0, \sigma)
    \end{equation}
    
    \item \textbf{Soft Target Network}: To update Q-values, a soft Q-function is used to estimate expected return with entropy:
    \begin{equation}
        \hat{Q}(x_t,a_t) = r(x_t,a_t) + \gamma E_{x_{t+1}\sim p} [V_{\bar{\psi}}(x_{t+1})] \quad 
    \end{equation}

    \item \textbf{Dual Critic}: Two Q-functions are employed as the critic, such that the minimum value is used to update the value and policy:
    \begin{equation}
        \resizebox{0.9\linewidth}{!}{$
        J_\pi(\phi)=\mathbb{E}_{x_t \sim \mathcal{D}}\left[D_{KL}\left(\pi'(\cdot|o_t) \,\middle\|\, \frac{\exp\left(Q^{\pi_{\text{old}}}(x_t,\cdot)\right)}{Z^{\pi_{\text{old}}}(x_t)}\right)\right]
        $}
    \end{equation}

    \item \textbf{Soft Target Network Updates}: The target network is updated using a smoothing coefficient ($\tau$) to improve stability:
    \begin{equation}
        \bar{\psi} \leftarrow \tau\psi + (1-\tau)\bar{\psi} \quad
    \end{equation}
\end{enumerate}

\subsubsection{Reward Function Formulation:} The adversarial goal is to maximise the mechanical stress (torque) on the target joint while remaining undetected by the anomaly detection system. The reward function is formulated as follows:

\scalebox{0.75}{
\begin{minipage}{\linewidth}
\begin{equation}
R(x_t, a_t, x_{t+1}) =
\begin{cases}
\tau_{\text{target}} \cdot (1 - P_{\text{anom}}), & \text{if waypoints are reachable} \\
-1, & \text{otherwise}
\end{cases}
\label{sec:reward-eq}
\end{equation}
\end{minipage}
}

where:
\begin{itemize}
    \item $\tau_{\text{target}}$ represents the total torque applied to the targeted robotic joint.
    \item $P_{\text{anom}}$ is the probability of detection assigned by the anomaly detection system.
    \item A penalty of $-1$ is assigned if the manipulated waypoints are deemed unreachable.
\end{itemize}

\subsubsection{Training Procedure}:

The adversarial agent is trained using experience replay with the following steps:

\begin{enumerate}
    \item Sample mini-batches from the replay buffer.
    \item Compute new target Q-values using the soft target network.
    \item Update soft value network using the function: 
    \begin{equation}
    \scalebox{0.7}{$J_V(\psi) = \mathbb{E}{x_t\sim D} \left[ \frac{1}{2} \left( V\psi(x_t) - \mathbb{E}_{a_t\sim \pi\varphi} [Q_{\theta_i}(x_t, a_t) - \log \pi_\varphi(a_t|x_t)] \right)^2 \right]$}
    \end{equation}
    \item Update the soft Q-function parameters using the target value network:
    \begin{equation}
    \scalebox{0.85}{$J_Q(\theta_i) = \mathbb{E}{(x_t, a_t)\sim D} \left[ \frac{1}{2} \left( Q_{\theta_i}(x_t, a_t) - \hat{Q}(x_t, a_t) \right)^2 \right] \quad$}
    \end{equation}
    \item Stochastic policy update using dual critic through the minimisation of KL divergence between the two critic networks.
    \item Update target network using smoothing. 
\end{enumerate}

By leveraging the POMDP framework, the adversarial strategy effectively learns stealthy manipulation techniques while maximizing wear-out impact in a highly dynamic industrial robotic environment. Additionally, during training the agent has access to manufacturer safety limits as the digital-twin model is known - at deployment the agent no longer has access to safety limits. However, the agent does have access to anomaly scores due to adversary access at Network L2 - exposing the Anomaly detection systems and data historian.

\section{Design \& Implementation}
\label{sec:Design and Implementation}

In this section, we describe the experimental setup considered to evaluate the proposed DRL-assisted wear-out attack against a DT-enabled industrial robotic system. 

\subsection{Simulation}\label{sec:simulation}
The experiments are conducted using a realistic physics engine - MuJoCo \footnote[2]{\url{https://mujoco.org/}} with an in-built simulation to the commercial specification of the industrial UR10e robot. MuJoCo is an advanced physics engine that enables fast, accurate simulation of articulated structures interacting with their environment. It is used for research and development in areas like robotics and biomechanics. We use the high-quality UR10e MuJoCo model from MuJoCo Menagerie \footnote[3]{See \url{https://github.com/google-deepmind/mujoco_menagerie} for UR10e MuJoCo model used}, as it represents a well-designed and accurate model of the real robotic arm, providing us with confidence that the simulated arm's operation will be as realistic as possible and our results valid. Unfortunately, while MuJoCo allows you to move each of the arm's joints under realistic physics rules, it does not provide any out-of-the-box capability of commanding the arm to reach specific waypoints. As such, we used the Robotics Toolbox for Python library \footnote{\url{https://petercorke.github.io/robotics-toolbox-python/}} to plan the trajectory that each of the joints should follow in order to reach the designated waypoints. The model's actuators were then used to execute the planned trajectory and move the arm's end effector to the intended waypoint coordinates.

\textbf{Wear-Out Measurement}
To measure wear-out and mechanical stress on the targeted joint, we use \textit{torque (Nm)}, as it is a good proxy for joint wear \cite{robot_joint_wear} and a quantifiable measure we can evaluate in our simulation. MuJoCo's main data structure (``mjData"), which holds the simulation's state, also keeps track of the torque applied to each of the joints \footnote{\url{https://github.com/google-deepmind/mujoco/issues/1095}}, allowing us to easily access that information and use it for the calculation of our reward (as explained in the DRL Environment). Other metrics also exist that can be used to approximate the wear-out of robotic joints such as joint vibrations and temperatures \cite{robot_joint_wear}, however, this kind of information is not easily accessible in a MuJoCo simulation.

\textbf{Target Joint}
In all of our experiments against UR10e, the targeted joint we try to stealthily wear out is ``Wrist 3", as this is one of the smallest joints of the arm and thus is designed to withstand the least torque\footnote{\url{https://www.universal-robots.com/articles/ur/robot-care-maintenance/max-joint-torques-cb3-and-e-series/}}. As such, an increase in the torque suffered by this joint is going to be more impactful than in other joints, resulting in quicker mechanical wear-out.
 
\subsection{DRL Agent:}\label{sec:rl_agent}
The \ac{DRL} agent is responsible for modelling the robotic arm's safe zones of an operation via the \ac{DT} and conducting intelligent wear-out attacks on the targeted joint. The agent's actions are modifications that get applied to the original waypoint coordinates the robotic arm reaches, with the goal of introducing subtle changes to the arm's trajectory and pose that induce additional strain on the targeted joint. This way, the targeted joint wears out sooner, thus increasing the owner's maintenance costs.  Here, we evaluate four \ac{DRL} algorithms, namely \ac{SAC} \cite{haarnoja2018softactorcriticoffpolicymaximum}, \ac{TD3} \cite{td3}, \ac{PPO} \cite{schulman2017proximalpolicyoptimizationalgorithms} and \ac{A2C} \cite{mnih2016asynchronousmethodsdeepreinforcement}, using their Stable-Baselines3\footnote{\url{https://stable-baselines3.readthedocs.io/en/master/}} implementations. The said algorithms are selected due to their popularity, high performance and compatibility with continuous action spaces such as ours. The main hyperparameters used for each algorithm can be found in Table \ref{table:alg_hyperparams}. 

\begin{table}[h]
\begin{tabular}{|p{2.3cm}|p{1.1cm}|p{1.1cm}|p{1cm}|p{0.9cm}|}
\hline
                                             & \textbf{SAC}     & \textbf{TD3}       & \textbf{PPO}    & \textbf{A2C}    \\ \hline
\textbf{Learning rate}                                & 0.0003  & 0.001     & 0.0003 & 0.0007 \\ \hline
\textbf{Buffer size}                                  & 1000000 & 1000000   & -      & -      \\ \hline
\textbf{Batch size}                                   & 256     & 256       & 64     & -      \\ \hline
\textbf{Tau}                                          & 0.005   & 0.005     & -      & -      \\ \hline
\textbf{Gamma}                                        & 0.99    & 0.99      & 0.99   & 0.99   \\ \hline
\textbf{Action noise}                                 & $\mathcal{N}(0,\,0.1)$    &  $\mathcal{N}(0,\,0.1)$ & -      & -      \\ \hline
\textbf{Entropy Regularisation Coefficient}                                 &  Auto   &  - & -      & -      \\ \hline
\textbf{Target policy noise}                          & -       & 0.2       & -      & -      \\ \hline
\textbf{Target noise clip}                            & -       & 0.5       & -      & -      \\ \hline
\textbf{Number of steps}                            & -       & -         & 128    & 5      \\ \hline
\textbf{Number of epochs}                             & -       & -         & 10     & -      \\ \hline
\textbf{Generalised Advantage Estimator lambda} & -       & -         & 0.95   & 1.0    \\ \hline
\textbf{Clip range}                                   & -       & -         & 0.2    & -      \\ \hline
\end{tabular}
\caption{Algorithms' Hyperparameters}
\label{table:alg_hyperparams}
\end{table}

\textbf{DRL Environment:}
A Gymnasium\footnote{\url{https://gymnasium.farama.org/index.html}} \ac{DRL} environment was developed that uses the aforementioned MuJoCo simulation to provide a space for the \ac{DRL} agent to interact, learn and be evaluated in. For the experiments conducted in this study, the robotic arm is configured to reach three different waypoints specified within the simulation environment. All of the original waypoints are reachable by the arm without additional strain on the joints. Furthermore, the arm is configured to begin each episode from a fully expanded horizontal home pose, move to each of the specified waypoints one by one, and finally return to its home pose.
The observation and action spaces were implemented to reflect the attacker's observation and action capabilities as defined in Section \ref{sec:ThreatModelling}.
The observations are standardised to have a mean of 0 and standard deviation of 1 using running statistics. Actions on are normalised in range [-1, 1] as far as the \ac{DRL} algorithm is concerned, however it is important to note that the actual modifications applied to the original waypoint coordinates are still in the range [-0.1, 0.1] . The scaling of observations and actions is very important in \ac{DRL} as it helps the Neural Networks used by the \ac{DRL} algorithms to have more stable and efficient training.

The agent's reward function is designed to maximise torque at the targeted joint while simultaneously minimising the likelihood of detection. The reward function is mathematically defined in Equation \ref{sec:reward-eq}.

Initially, the environment checks whether the adversarially modified waypoints are reachable by the arm. If they are deemed unreachable, or if the action performed exceeds the safety limits specified by the manufacturer, the agent is penalised with a reward of \(-1\) to discourage infeasible actions. If the waypoints are reachable, the MuJoCo simulation runs, calculating the total torque applied to the targeted joint and assessing the anomaly probability of the executed arm trajectory using an anomaly detection system that is detailed later in the paper.

The integration of the anomaly detection system in the reward function guides the agent to execute \textit{stealthy} attacks. Similar to observations and actions, rewards are scaled to maintain a standard deviation of 1 while preserving their original means. This ensures consistent reward magnitude, favouring effective training,  without altering the reward sign, which could negatively impact the agent's learning process. 

\subsection{Anomaly Detection System}\label{sec:anomalyDetector}
A comprehensive anomaly detection system capable of recognising abnormal robotic arm trajectories is used to assess the DRL agent’s stealthiness and support its training. The system consists of an ensemble of four Autoencoder-based anomaly detectors (\ac{LSTM}, \ac{CNN}, \ac{ResNet}, and \ac{GRU}), each using joint state data (positions, velocities, accelerations) at each time step to predict the probability of an anomaly. Each detector’s prediction is weighted by its test F1-score to produce an overall anomaly probability, which is incorporated into the reward function to guide the agent toward taking \textit{undetectable} malicious actions.

The detectors are trained on sequences of joint states collected under normal operation (i.e., following the original waypoints) and, for time-enabled environments, the corresponding timestamps. For the training, validation, and testing of these anomaly detection systems a 76.4\%/13.6\%/10\% split has been used over a dataset of 400,000 coordinate and joint position values collected over 16,000 noise inclusive trajectories. To emulate realistic behaviour, normal trajectories are captured with ±0.1 mm noise added to each waypoint, consistent with the UR10e specifications\footnote{\url{https://www.universal-robots.com/manuals/EN/HTML/SW5_19/Content/prod-usr-man/complianceUR10e/H_g5_sections/appendix_g5/tech_spec_sheet.htm}}. This allows them to learn a latent representation of normal arm behaviour. They operate by encoding normal inputs into a compressed representation and reconstructing them, similar to Kalman-filter residual analysis. Since they are trained solely on normal data with Gaussian noise, they reconstruct such data with minimal error. During inference, if a reconstruction exceeds a predefined threshold, it is flagged as anomalous. Thresholds are selected for each detector by training on normal and abnormal sequences and choosing the value that maximises F1-score. In known environments, the safety limit also serves as a rule-based detector: any action that breaches it is marked as anomalous. We provide a more in-depth discussion on IDS performance in Appendix \ref{app:IDS}.

\begin{table}[h]
\begin{tabular}{|P{1.8cm}|P{3.5cm}|P{1.5cm}|}
\hline
\textbf{Intrusion Detection Model} & \textbf{Architecture}                                                                                     & \textbf{F1-Score (\%)} \\ \hline
LSTM AE       & \ac{LSTM} layers, Fully Connected Layers  & 96.2  \\ \hline
LSTM          & \ac{LSTM} Layer, Fully Connected Layer & 100.0\\ \hline
CNN AE        & Convolutional 1D layers, \ac{ReLU} layers & 97.4  \\ \hline
ResNet AE     & Convolutional 1D layers, \ac{ReLU} layers, Batch Normalisation 1D layers, Max Pooling 1D layers, Residual Connections   & 99.8 \\ \hline
GRU AE        & \ac{GRU} Layers, Fully Connected Layers&  98.3 \\ \hline
\end{tabular}
\caption{Anomaly Detectors' Architectures and F1-Scores}
\label{table:anom_detectors}
\end{table}

\section{Experiments \& Results}
\label{sec:Experiments and Results}
In this section, we detail the experiments conducted to determine the most appropriate \ac{DRL} algorithm for performing the desired attack. Additionally, we rigorously evaluate its performance across different scenarios and the effects of key hyperparameters. It is important to note that while the IDS mechanisms have been tested in a time-enabled environment (which provides resilience against replay attacks), the time difference between normal and manipulated behaviour is negligible. As a result, the experiments are conducted in the default environment, and temporal factors are not considered. 

\subsection{Algorithm Benchmarking}
In order to determine the most suitable \ac{DRL} algorithm for conducting the intended attack, the algorithms mentioned in Section \ref{sec:rl_agent} are compared and contrasted based on their performance against the \textit{base case}. The base case represents the scenario where the attacker has access to all relevant environmental observations (i.e., joint positions, velocities, and accelerations), targets the joint "Wrist 3" of the robotic arm, and the agent trains for 10,000 episodes. Benchmarking the algorithms on the same scenario enables us to objectively and fairly determine which one is more suitable for the formulated problem. Each algorithm has been tested for a period of 100 episodes after training to further evaluate performance.

\begin{figure}[ht!]
\centering
\begin{subfigure}{0.5\textwidth}
  \centering
  \includegraphics[width=0.85\linewidth]{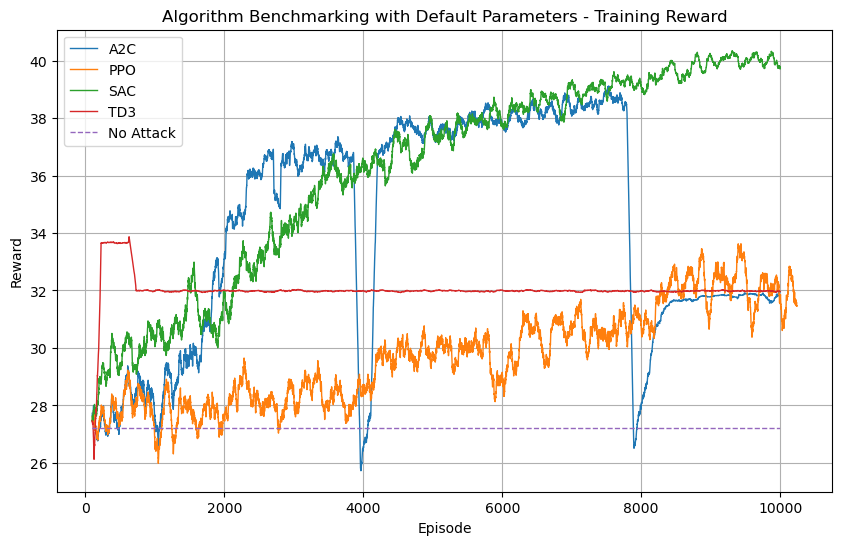}
  \caption{}
  \label{fig:alg_bench_train}
\end{subfigure}
\begin{subfigure}{0.5\textwidth}
  \centering
  \includegraphics[width=0.85\linewidth]{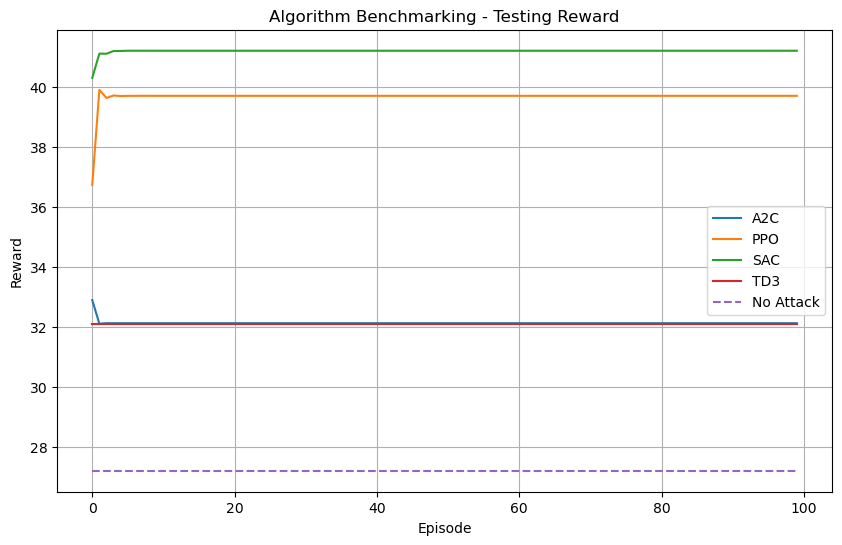}
  \caption{}
  \label{fig:alg_bench_test}
\end{subfigure}
\caption{Algorithm Benchmarking performance during (a) training  and  (b) testing}
\label{fig:alg_bench}
\end{figure}

The training and testing performance of each algorithm is plotted in Figure \ref{fig:alg_bench}, and further benchmarking metrics can be found in Table \ref{table:alg_benchmark_metrics}.

As shown in Figure \ref{fig:alg_bench}, \ac{SAC} performs the most consistently across training and testing. During training, all algorithms manage to obtain better performance than the baseline of ``No Attack", which represents the normal operation of the robotic arm. During testing, \ac{SAC}, and \ac{PPO} achieve significant improvements over the baseline, whereas \ac{TD3} and \ac{A2C} demonstrate significantly weaker overall performance.

Comparing the on-policy algorithms (\ac{PPO}, and \ac{A2C}): \ac{PPO} demonstrates a consistent performance due to its clipping mechanism that prevents large changes to its policy; however, this causes the policy to be restrictive in its exploration. \ac{A2C} presents a stronger, albeit much more unstable performance as it does not incorporate clipping, which allows it to better adapt to the gymnasium environment and the continuous action space.

Out of the off-policy methods (\ac{SAC}, \ac{TD3}: \ac{SAC} demonstrates consistent improvements over the training period while remaining stable, whereas, \ac{TD3} shows strong early improvements before converging at a sub-optimal policy. Throughout training, \ac{SAC}'s performance can largely be attributed to its inherent stochastic nature, as its entropy regularisation coefficient encourages exploration, introduces adaptability, and helps avoid local optima, thereby allowing convergence to more optimal policies. \ac{TD3}, on the other hand, is a deterministic policy method, leading to poorer exploration capabilities - hence the instability during training. 

As shown in Table \ref{table:alg_benchmark_metrics}, SAC attains the largest AUC, with A2C following and TD3 and PPO trailing due to reduced learning. This reflects the clear sample-efficiency advantage of the off-policy methods, whose replay buffers enable faster and more stable cumulative learning than their on-policy counterparts\footnote[2]{For a detailed comparison between off-policy and on-policy DRL algorithms see \cite{sutton2018reinforcement}.}. Lastly, it should be noted that with a substantially larger number of training episodes, less sample-efficient on-policy algorithms could attain performance similar to \ac{SAC}.

Following the benchmarking experiments, additional environment configuration experiments have been conducted to determine the optimal values for both the agent's action and observation spaces. The results of these experiments are presented in  Section\ref{sec:action-space-exp} and Section \ref{sec:obs-space-exp}.

\begin{tcolorbox}
\textbf{Takeaways:} \ac{SAC} and \ac{A2C} are very effective in conducting a stealthy wear-out attack in this environment. \ac{SAC}'s superior sample efficiency and consistent performance during both training and testing make it the most appropriate algorithm overall for this study and is therefore used for all subsequent experiments. It should be noted, however, that PPO and A2C remain strong candidates and were only excluded due to instability observed during training.  
\end{tcolorbox}

\begin{table}[ht!]
\begin{tabular}{|p{2cm}|p{1.1cm}|p{1.1cm}|p{1.1cm}|p{1.1cm}|}
\hline
& \textbf{SAC} & \textbf{TD3} & \textbf{PPO} & \textbf{A2C} \\ \hline
\textbf{Mean Training Reward}      & 38.90 & 28.80 & 31.79 & 40.74\\ \hline
\textbf{Mean Testing Reward}       & 36.13 & 32.00 & 29.72 & 34.10 \\ \hline
\textbf{Area Under Learning Curve} &  
361281.49 & 319951.22 & 304346.14 & 340875.76\\ \hline
\end{tabular}
\caption{Algorithms' Benchmark Metrics}
\label{table:alg_benchmark_metrics}
\end{table}

\subsection{Exploration vs Exploitation}

The exploration–exploitation trade-off is a well-known challenge in \ac{DRL} \cite{wang2019explorationversusexploitationreinforcement}, where the agent must balance trying new actions (exploration) with selecting the best-known action (exploitation). If the agent does not explore sufficiently, it may converge prematurely to a suboptimal policy, whereas excessive exploration can slow convergence and prevent the agent from stabilising an effective strategy.

\ac{SAC} manages this trade-off through an entropy regularisation coefficient in its objective function \cite{SAC_Entr_Reg}. This coefficient determines how strongly the agent is encouraged to maintain stochasticity in its policy. Higher values promote greater exploration, while lower values place more emphasis on exploiting learned behaviour.

In this experiment, we evaluate a range of entropy coefficients: 0.1, 0.3, 0.5, 0.7, 0.9, and “auto”, to analyse their impact on performance. When “auto” is used, the entropy coefficient is learned dynamically during training.

\begin{figure}
\centering
\begin{subfigure}{0.5\textwidth}
  \centering
  \includegraphics[width=0.85\linewidth]{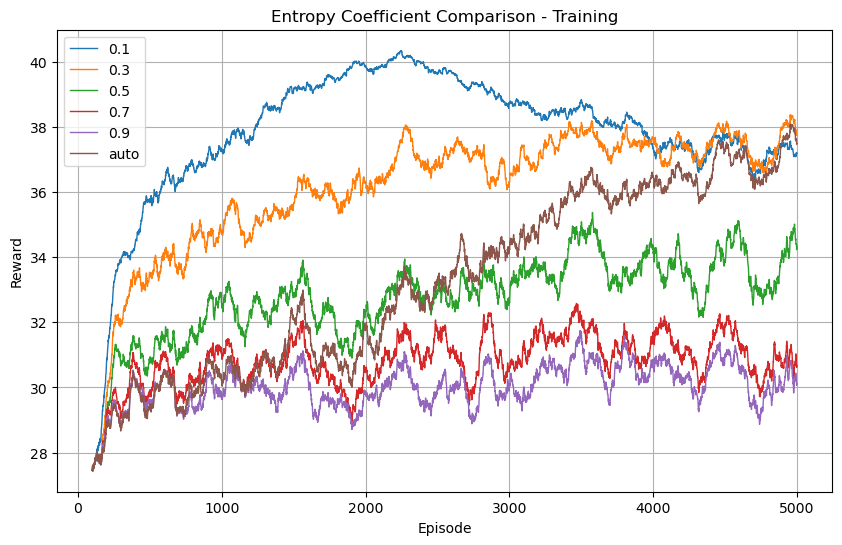}
  \caption{}
  \label{fig:explor_exploit_train}
\end{subfigure}
\begin{subfigure}{0.5\textwidth}
  \centering
  \includegraphics[width=0.85\linewidth]{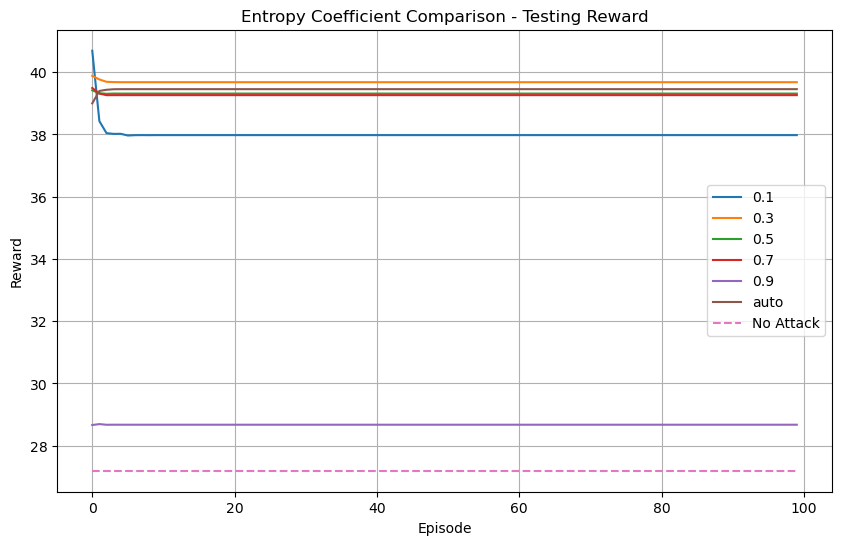}
  \caption{}
  \label{fig:explor_exploit_test}
\end{subfigure}
\caption{Entropy Regularisation Coefficient comparison during  (a) training and (b) testing. }
\label{fig:explor_exploit}
\end{figure}

Figure \ref{fig:explor_exploit_train} shows that the lowest coefficient (0.1) consistently yields the strongest training performance, followed by 0.3. These settings enable the agent to exploit effective behaviours early while maintaining enough stochasticity to avoid premature convergence. The remaining coefficients, particularly 0.7 and 0.9, perform substantially worse, demonstrating that excessive exploration prevents the policy from stabilising and leads to noticeably lower cumulative returns. The “auto’’ configuration performs comparably to the mid-range coefficients but does not surpass the best fixed setting.

Testing results in Figure \ref{fig:explor_exploit_test} reveal that, apart from the highest coefficient (0.9), all settings converge to similar evaluation performance, with 0.3 and 0.5 achieving the strongest results. This indicates that while low-entropy configurations accelerate learning, moderate exploration can yield slightly more stable behaviour at deployment. The poor performance of the 0.9 coefficient in both training and testing highlights the detrimental effect of persistent over-exploration.

Overall, the results show that lower entropy coefficients provide the most favourable learning dynamics, while moderate coefficients offer competitive evaluation performance. Excessively large coefficients hinder both convergence and final reward, highlighting the importance of carefully balancing exploration and exploitation when training adversarial SAC agents.

\begin{tcolorbox}
\textbf{Key Takeaways:} While exploration is necessary for diverse experience and avoiding local optima, excessive entropy reduces the ability to learn effectively. Therefore, it must be applied carefully to ensure effective convergence toward a near-optimal solution. Lower fixed entropy coefficients provide the most reliable exploration–exploitation balance and are used for all subsequent experiments - in this case, 0.3 is used.
\end{tcolorbox}

\subsection{Action Space}\label{sec:action-space-exp}

The action space determines the scale of perturbations the adversarial agent can inject into the control signal, thereby shaping both its learning dynamics and its eventual influence on the robotic policy. To assess how perturbation magnitude affects performance, we evaluate three ranges $[-0.1,0.1]$, $[-0.01,0.01]$, and $[-0.001,0.001]$, to span three orders of magnitude.

As shown in Fig. \ref{fig:action-space-training}, the two larger ranges, $[-0.1,0.1]$ and $[-0.01,0.01]$, enable higher rewards during training, with the widest range yielding the fastest and strongest improvement. In contrast, the smallest range produces almost no learning signal and remains close to the no-attack baseline throughout, indicating that perturbations of this scale are too weak to meaningfully alter the victim’s trajectory during training.

However, the evaluation results in Fig. \ref{fig:action-space-testing} show that the intermediate range, $[-0.01,0.01]$, achieves the strongest performance at test time. While $[-0.1,0.1]$ still performs well, its impact is consistently weaker than that of the intermediate range, and the smallest range again fails to surpass the no-attack baseline. These findings indicate that although broad action ranges accelerate learning by enabling the agent to quickly discover impactful manipulations, moderately sized perturbations provide a more effective balance between influence and stability during deployment

\begin{tcolorbox}
\textbf{Key Takeaways:} Larger action spaces improve exploration during training, but excessively large adjustments reduce stealth and reduce the agent's ability to generalise to evaluation conditions. Conversely, very restrictive ranges limit the agent’s ability to influence the system. The intermediate action range of $[-0.01, 0.01]$ therefore offers the most effective trade-off, providing sufficient manipulative capability while remaining subtle enough to avoid detection, meaning this will be the action range for further experimentation.
\end{tcolorbox}

\begin{figure}[ht!]
    \centering
    \includegraphics[width=0.85\linewidth]{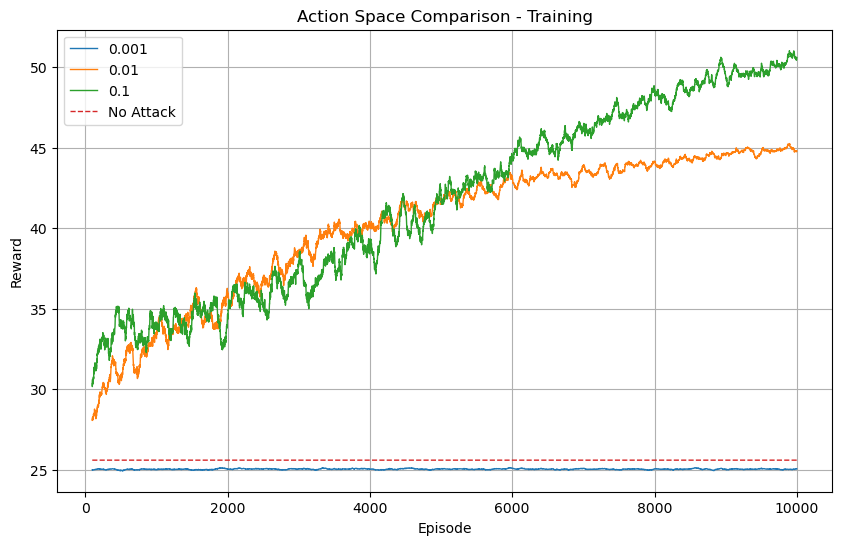}
    \caption{Action Space Experimentation - Training Rewards}
    \label{fig:action-space-training}
\end{figure}
\begin{figure}
    \centering
    \includegraphics[width=0.85\linewidth]{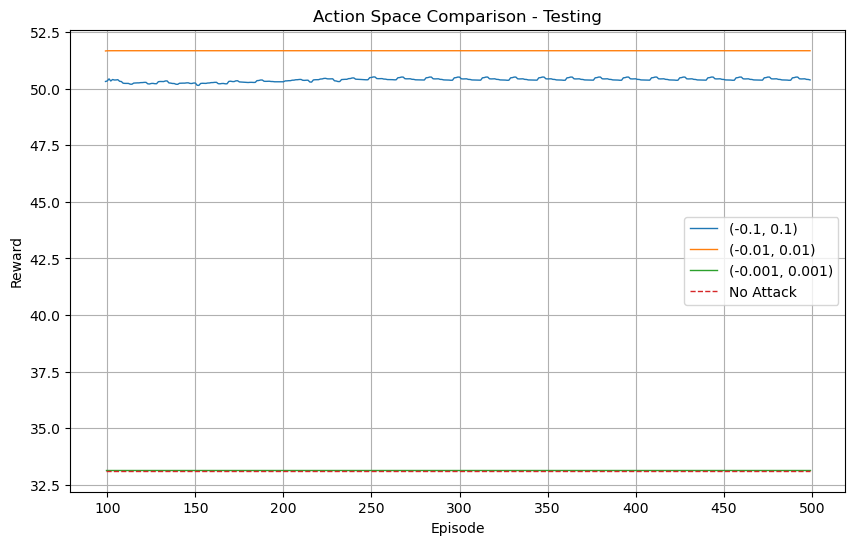}
    \caption{Action Space Experimentation - Testing Rewards}
    \label{fig:action-space-testing}
\end{figure}

\subsection{Observation Space \& Rewards Relation}\label{sec:obs-space-exp}

The amount and type of data the adversarial agent has access to directly impact its ability to conduct the attack and its effectiveness. As such, in this set of sub-experiments, we assess the adversarial agent's effectiveness across varying knowledge spectra. For each sub-experiment, the agent has access only to a subset of the original observation space (joint positions, joint velocities, and joint accelerations), following a grey-box threat model. This set of sub-experiments can also help determine which types of observations are most important for successfully conducting a stealthy wear-out attack.

The observation space used for each sub-experiment is:
\begin{itemize}
    \item Sub-experiment 1: Sub-experiment 4: Full joint information of half of the joints only (namely ``Shoulder", ``Wrist 1" and ``Wrist 3" joints)
    \item Sub-experiment 2: Sub-experiment 5: Full joint information of all joints (six joints - base joint is excluded for the Panda model)
    \item Sub-experiment 3: Joint position only
    \item Sub-experiment 4: Joint velocity only
    \item Sub-experiment 5: Joint acceleration only
\end{itemize}

Sub-experiments 1–2 were conducted first to determine how many known joints are necessary for the agent to perform an effective attack. Table \ref{table:obs_space_metrics} show that the training performance across all observation configurations is nearly identical, indicating that increasing the number of joints or adding additional features provides no clear learning advantage. The testing results in Table \ref{table:obs_space_metrics} reinforce this conclusion: all configurations yield almost the same reward, substantially outperforming the no-attack baseline and demonstrating that the agent can execute a successful attack even with minimal information.

Given this consistency, sub-experiments 3–5 restrict the observation space to a single joint. These results show only small differences between using position, velocity, or acceleration, with joint position performing marginally better. Overall, the agent performs well with very limited information, and the observation space has little impact on effectiveness. If anything, simpler observation spaces lead to slightly improved performance by reducing unnecessary complexity without diminishing the agent’s ability to disrupt the system.

\begin{tcolorbox}
    \textbf{Key Takeaways}: Although different observation spaces produce only minor variations in performance, the agent performs consistently well across all configurations, achieving rewards of approximately 39 in each case. Joint position emerges as the least influential feature, with simpler, alternating joint (3 joints of the 6) observation spaces yielding the strongest results. Therefore, to determine how well the agent performs with the least information, the remaining experiments will use single joint position values.
\end{tcolorbox}

\begin{table}[h]
\begin{tabular}{|p{2.5cm}|P{2cm}|P{2cm}|}
\hline
\textbf{} & \textbf{Mean Training Reward} & \textbf{Mean Testing Reward} \\ \hline
\textbf{Full Information}          & 33.10 & 39.46 \\ \hline
\textbf{Partial Information}       & 33.13 & 39.77 \\ \hline
\textbf{Single Joint Information}  & 33.28 & 39.39 \\ \hline
\textbf{Single Joint Acceleration} & 33.20 & 39.63 \\ \hline
\textbf{Single Joint Velocity}     & 33.28 & 39.41 \\ \hline
\textbf{Single Joint Position}     & 33.25 & 39.45 \\ \hline
\end{tabular}
\caption{Observation Space and Rewards relation comparison}
\label{table:obs_space_metrics}
\end{table}

\subsection{Wear-Out Impact}

In order to facilitate the comprehension of a trained agent's attack impact against the victim's robotic arm, we compare it against the baseline ``No Attack" case with respect to the total torque suffered by the target joint over 1000 episodes. For this and all subsequent experiments, the trained SAC agent with  ``auto" entropy regularisation coefficient and access to only the target joint's position information (as far as joint states are concerned) is used, as this is the algorithm setup that achieves the best testing performance according to the previous experiments (See section \ref{sec:action-space-exp} for the configuration experiment results).

In the ``No Attack" case, the mean torque at the target joint is 33.12 Nm, whereas with our trained agent, it increases to 51.15 Nm. This shows that the attack successfully increases the torque applied to the target joint by approximately 54.44\% compared to its normal operation.

Since the lifespan of a material decreases exponentially with the increase in stress \cite{japp2014fatigue}, the reduction factor can be computed as $(\frac{\sigma_{attack}}{\sigma_{normal}})^k$, with $\sigma$ representing stress levels and $k$ the material's fatigue exponent (usually ranges from 6 to 10 for aluminium materials \cite{stephens2000metal} used in UR10e\footnote[2]{\url{https://www.universal-robots.com/media/50880/ur10_bz.pdf}}). The reduction factor can be calculated as (where \ref{eq:red-fac-low} and \ref{eq:red-fac-high} show the lower and upper bounds respectively):
\begin{equation}\label{eq:red-fac-low}
    \text{Reduction Factor} = \left(\frac{51.15}{33.12}\right)^6 \approx 1.544^6 \approx 13.55
\end{equation}
\begin{equation}\label{eq:red-fac-high}
    \text{Reduction Factor}=\left(\frac{51.15}{33.12}\right)^{10}\approx1.54^{10}\approx 77.00
\end{equation}
Assuming that joint ``Wrist 3" has a lifespan of 35000 hours \footnote[3]{\url{https://www.universal-robots.com/media/8641/ur_brochure_gb.pdf}}, the lifespan of the joint if it is under constant attack by our agent is:
\begin{equation}
   L_{attack}=\frac{L_{normal}}{Reduction\ Factor}=\frac{35000}{13.55}\approx 2583
\end{equation}
\begin{equation}
       L_{attack}=\frac{L_{normal}}{Reduction\ Factor}=\frac{35000}{77.0}\approx 454.5
\end{equation}

Based on the above calculation, if the target joint operates continuously under the agent's attack torque, its lifespan is drastically reduced to between approximately 454.4-2583 hours depending on the material fatigue.

\begin{tcolorbox}
\textbf{Takeaways:} 
The trained agent increases the torque induced on the target joint by approximately 54.44\% compared to the robotic arm's normal operation. This significantly decreases the component's lifespan from 35000 hours to between 454.5 and 2583 hours depending on the material fatigue limits, thus increasing the maintenance costs.
\end{tcolorbox}

\subsection{Undetectability}

One of the key desired characteristics of the adversarial agent is that it is able to conduct the attack in a stealthy manner, ie remain undetected by the security mechanisms in place within the network. In this experiment, we test the agent's ability to perform manipulations to the control signal while remaining undetectable by the \ac{IDS} specified in Section \ref{sec:Design and Implementation}. For the purpose of this experiment, the agent is considered to be detectable if the anomaly score exceeds $50 \%$ - ie it is more likely that the value is anomalous than not. The attack experiment has been run for 1000 episodes  in order to gather information from a sufficient number of interactions between the agent and the anomaly detection system.

In this experiment, the adversarial agent achieved an average anomaly score of  $\approx 20.63 \%$, whereas, the baseline produces an average anomaly score of $\approx17.91\%$ due to standard process noise defined in the UR10e specification \cite{Universal-Robots}.  These results demonstrate that the perturbations introduced by the adversary are perceived as slightly more anomalous than the standard process noise. This means that the agent has learned to make minor adjustments that introduce higher mechanical stress to the targeted joint while effectively blending in with standard operating behaviour.  In addition to showcasing the agent's ability to perform the attack in a stealthy manner, the low anomaly scores produced by waypoints introduced by the agent's actions indicate that the agent can exploit deficiencies within the ensemble \ac{IDS} deployed in the simulation environment. When this is combined with the results from the "Wear-Out-Impact" experiments, the agent demonstrates that it can achieve two clear goals: cause stress-related wear-out while remaining undetectable within the network, making the attack particularly dangerous.

\begin{tcolorbox}
{\textbf{Key Takeaways}: The agent is able to achieve an average anomaly score of $\approx 20.63\%$, producing a slightly higher anomaly probability compared to the standard operating behaviour of the robotic arm established in the baseline ($\approx 17.91 \%$ due to process noise). This demonstrates that agent is able to maintain undetectability while exploiting deficiencies within the deployed \ac{IDS}}.
\end{tcolorbox}

\subsection{Naive Attacks Comparison}
In order to evaluate the performance of the \ac{SAC} adversarial agent against both the current, more naive methods, we have performed experiments against two naive methods: constant adjustment, and adding random values in the range $[-0.01, 0.01]$. This experiment was run for 1000 episodes in a similar manner to prior experiments, in order to collect enough interactions between the adversary and the anomaly detection systems. The first attack is a "constant" attack, such that a constant value (in this case 0.07) is introduced to the original waypoint in order to adjust the standard behaviour, the second is a "random" attack, where small random modifications in the range $[-0.01,0.01]$ are sampled from a uniform normal distribution to replicate the action space used by the agent while focusing on a stochastic approach. The performance of these attacks is compared to the SAC agent, and an established baseline in Table \ref{table:naive_attacks_vs_sac}

The outcomes of this experiment demonstrate that both of the naive methods perform above the baseline by $\approx 8$ points; however, both of these methods lead to a decrease in the mechanical stress applied to the targeted joint, which has the effect of increasing the lifespan of the joint as opposed to increasing stress to reduce the lifespan. When considering this result alongside the anomaly probability, it becomes evident that the naive attack methods perform very minor adjustments, so much so that the \ac{IDS} considers the change in behaviour as less noisy than the standard process noise, leading to a minimal anomaly score which boosts the reward received despite failing to accomplish the attack goals. Comparing these results to the performance achieved by the SAC agent, we can see that the agent drastically outperforms both of the naive models, both with respect to the average reward received and the average torque applied to the targeted joint. 

\begin{tcolorbox}
    \textbf{Key Takeaways:} The adversarial agent demonstrates a considerable improvement over the implementation of the naive attack methods. Its strategic modification of the waypoints leads to a notable increase in stress applied at the target joint, while maintaining stealth within the network, and succeeds in both attack goals, whereas the naive methods only succeed at remaining undetectable over a 1000 episode period.
\end{tcolorbox}

\begin{table}[h]
\begin{tabular}{|P{1.5cm} |P{1.5cm}|P{1.3cm}|P{1.3cm}|P{0.7cm}|}
\hline
& \textbf{Normal Operation} & \textbf{Constant Attack (0.07)} & \textbf{Random Attack} & \textbf{SAC} \\ \hline
\textbf{Mean Reward}                    & 27.20 & 32.28 & 33.13 & 40.60  \\ \hline
\textbf{Mean Target Joint Torque (Nm)}  & 33.12 & 32.28 & 33.13 & 51.15\\ \hline
\textbf{Mean Anomaly Probability (\%)}  & 17.91 & 1.31e-05 & 1.48e-05 & 20.63\\ \hline
\end{tabular}
\caption{Naive Attacks vs SAC Performance Comparison}
\label{table:naive_attacks_vs_sac}
\end{table}

\subsection{Out of Context}

One of the key advantages of DRL agents is their ability to operate in unforeseen environments and adapt to new situations. In this experiment, we evaluate the adversarial agent’s generalisability to unseen robotic arms on two new models: first, the UR5e (made by the same manufacturer as the UR10e) to assess performance in a semi-similar environment, and second, the Franka Panda, which introduces a substantially different configuration. The Franka Panda also differs in its joint structure, meaning that the target joint is “Joint 7” rather than “Wrist 3” in the UR models\footnote{Franka Panda’s “Joint 7” is the arm’s final joint, analogous to “Wrist 3” in UR10e.}. Additionally, the tolerance used in the inverse-kinematics solver inflates torque values for the Panda model. Although increasing this tolerance mitigates the effect, it also prevents the arm from moving, as the permissible error becomes larger than the adversarial perturbations.

For consistency, we retain the original tolerance from the UR experiments and focus primarily on the agent’s behavioural trends in the Panda environment rather than the absolute torque values. Two sub-experiments are used to measure adaptability: the first deploys a “fresh” SAC agent in each new environment, and the second applies a pre-trained agent with a 2500-episode transfer-learning period (a quarter of the original training duration). For the Panda model, the agent is pre-trained on both UR10e and UR5e.

Table \ref{table:out_of_context_comparison} summarises these results. In the semi-known UR5e environment, both SAC agents increase the torque at the target joint relative to normal operation (8.07 Nm). The fresh agent produces a modest increase (8.92 Nm) with a small rise in anomaly probability, whereas the transfer-learning agent produces a substantially larger torque (17.66 Nm) but at a markedly higher anomaly probability (40.34\%). This suggests that while the agent accounts for anomaly likelihood when optimising reward, it ultimately prioritises force application when it can leverage prior knowledge. Depending on the adversarial goal, this may be beneficial, though increasing the weight of the anomaly term could encourage stealthier behaviour. Overall, the UR5e results show that the agent can adapt effectively to a similar environment and still achieve its primary attack objectives.

For the Panda model, the fresh SAC agent again increases the mean torque relative to normal operation, but the transfer-learning configuration does not lead to further improvement, instead failing to outperform the fresh baseline. This is likely due to the significant differences between the UR and Panda kinematic structures: while the UR10e and UR5e share characteristics that allow for beneficial transfer, the Panda model diverges substantially, limiting generalisation and in some cases hindering adaptation.

Finally, because the pre-trained agent relies only on the target joint’s positional information, it avoids any incompatibility in observation-space dimensionality despite the change in degrees of freedom. This requirement for only the target joint’s state allows the agent to operate across multiple robotic platforms with differing joint counts.

\begin{tcolorbox}
\textbf{Key Takeaways:} The pre-trained \ac{SAC} agent adapts effectively to environments that share structural similarities, producing greater torque increases than a freshly trained agent under. Its use of only the target joint’s position enables straightforward transfer across robotic arms with different degrees of freedom. However, in dissimilar environments such prior knowledge becomes detrimental, as behaviours learned on UR models fail to translate effectively and ultimately limit the agent’s performance.
\end{tcolorbox}

\begin{table}[h]
\begin{tabular}{|P{2.5cm}|P{1.5cm}|P{1cm}|P{1.5cm}|}
\hline
\multicolumn{4}{|c|}{\textbf{UR5e Results}} \\ \hline
& \textbf{Normal Operation} & \textbf{Fresh SAC} & \textbf{Transfer Learning SAC} \\ \hline
\textbf{Mean Reward}                   & 6.68 & 9.18 & 10.53  \\ \hline
\textbf{Mean Target Joint Torque (Nm)} & 8.07 & 8.92 & 17.66 \\ \hline
\textbf{Mean Anomaly Probability (\%)} & 17.20 & 20.15 & 40.34 \\ \hline

\multicolumn{4}{|c|}{\textbf{Franka Panda Results}} \\ \hline
& \textbf{Normal Operation} & \textbf{Fresh SAC} & \textbf{Transfer Learning SAC} \\ \hline
\textbf{Mean Reward}                   & 448.49  & 451.66  &  \\ \hline
\textbf{Mean Target Joint Torque (Nm)} & 1118.96 & 1126.81 &   \\ \hline
\textbf{Mean Anomaly Probability (\%)} & 59.91  & 59.91 & 59.91 \\ \hline
\end{tabular}
\caption{Normal Operation vs Fresh SAC vs Transfer Learning SAC Performance on UR5e and Franka Panda Environments}
\label{table:out_of_context_comparison}
\end{table}

\section{Discussion}\label{sec:Disc_FutWork}
Our study shows that a DRL-assisted adversary can effectively execute a stealthy wear-out attack on a DT-enabled industrial robotic system by subtly modifying the control signals transmitted from the \ac{VE} to the \ac{PE}. Unlike conventional attacks on DT infrastructures~\cite{secFrameworkDTCloud, prototypingIntrusionDetection, akbarian2020, balta2024}, our approach leverages DRL to develop an intelligent adversary that adapts to its environment in real time. This enables the adversary to execute persistent, long-term attacks while remaining undetected by existing security mechanisms, eliminating the need for manual tuning to balance impact and stealth. Our adversarial agent follows a strategic ``low \& slow'' attack methodology, carefully manipulating control signals to gradually increase the torque on the targeted joint by approximately 54\% — while maintaining a low anomaly profile. This ensures the attack's persistence by avoiding immediate detection and accelerating mechanical wear subtly over an extended period.

\subsection{DRL Algorithm and Agent Performance}
A key factor in the success of this stealthy adversarial approach is the choice of an appropriate DRL algorithm. Our benchmarking demonstrated that SAC, overall, yields the most effective and sample-efficient adversarial policy for stealthy wear-out attacks in our experimental setup. It also showed that off-policy DRL algorithms (\ac{TD3} and \ac{SAC}), generally demonstrate superior sample efficiency compared to on-policy DRL algorithms (\ac{PPO} and \ac{A2C}). This advantage of off-policy methods is attributed to their ability to store and reuse past experiences via replay buffers, which accelerates convergence and enhances learning efficiency. However, despite TD3's high sample efficiency, its deterministic nature hindered its convergence to an optimal policy, whereas SAC's stochastic approach encouraged the exploration of more beneficial policies. 

A useful insight gained from this research is the role of the entropy regularisation coefficient in balancing the exploration-exploitation trade-off for \ac{SAC} algorithm. Constrained or automatically adjustable entropy regularisation coefficients help the agent gather sufficient diverse experiences to escape local optima without excessively slowing convergence. The automatically adjustable entropy coefficient appeared to be the most suitable for balancing exploration and exploitation in this problem, yielding the highest performance. Furthermore, it is noteworthy that the trained \ac{DRL} agent can conduct a highly effective attack with minimal knowledge (e.g., only the position data of the targeted joint). This suggests that even the slightest amount of information can be sufficient for an intelligent agent to cause significant damage to the target system. Additionally, the DRL-assisted adversary's strategic actions were shown to lead to a considerably more effective attack compared to naive constant or random actions.

Another notable advantage of the adversarial agent trained is that it can discover and exploit weaknesses in the anomaly detection defence system used by the \ac{DT}. Specifically, it exploits deficiencies in the anomaly detection system, allowing it to construct damaging waypoints that are misclassified as less anomalous than the original waypoints. Moreover, the trained agent can adapt to new security mechanisms with which it has no prior experience. The pre-trained DRL agent was also shown to be generalisable and easily adaptable to new environments, regardless of the victim arm's degrees of freedom, and to effectively damage a completely different robotic arm with only a limited number of fine-tuning episodes.

\subsection{Task Complexity and Testbed}

The agent is trained and evaluated on navigation between three predefined waypoints, this is simpler than standard industrial operations. We argue, however, that the attack surface our adversary exploits is invariant to task complexity: ICS operation fundamentally consists of state-to-state transitions mediated by waypoint or set point commands, and our adversary manipulates commands within this framework. The control-loop primitives exercised in our setup—inverse kinematics resolution, trajectory planning, per-axis waypoint dispatch—are identical to those used in richer tasks such as multi-stage assembly or pick-and-place Extension to more complex tasks is reliant on agent training to handle the required operations. Our transfer-learning results in §5.8 support this: a pre-trained SAC adversary adapts to a structurally similar arm (UR5e) within a quarter of the original training time, while structurally dissimilar platforms require larger retraining phases. We expect similar behaviour across more varied trajectories, which will require proportionally more episodes to converge on an optimal attack, but the underlying mechanics remain unchanged. Training is performed entirely offline against a replicated DT model, and the resulting policy adds no network traffic at deployment, suggesting the main barrier for mounting this attack is access to the VE-to-PE communication rather than computational power.

We use MuJoCo as it supports extensibility across robotic platforms and reproducibility of the adversarial dynamics studied. Simulation allows other researchers to replicate our results, benchmark alternative defences against the same adversary, and adapt the methodology to new platforms. Several aspects of real-world deployment lie outside the scope of the present simulation. Extended operation under sustained adversarial torque would engage wear mechanisms beyond the stealthy wear-out employed, including more severe mechanical degradation, bearing fatigue, and specific failure modes, each of which would add to a more precise characterisation of component lifespan. Additionally a real-world deployment adds environmental considerations such as humidity, environmental and physical temperature, and power supply fluctuations - which may affect agent performance due to action trajectories, potentially leading to abnormal mechanical states.

\subsection{Defence Mechanisms}

The main contribution of this paper is offensive: a novel DRL-driven wear-out attack, and the public release of the associated code, datasets, and trained policies. By disclosing this material, we aim to enable defenders to harden existing IDS mechanisms against the class of stealthy, adaptive adversaries we describe. Beyond direct hardening, however, our framework is well-suited to a more defensive strategy: using the offensive policy as the training partner for a learning-based defender.

An extension of our framework is to treat the defender as an agent embedded in the DT pipeline, observing joint telemetry and selecting defensive actions. The defender's reward captures the objective of detecting manipulated trajectories while minimising false positives on legitimate operation. The defender can be trained against the offensive policies generated by our framework rather than pre-recorded attacks. This exposes the agent to precise low-and-slow manipulations that the reconstruction-based detectors fail to detect. Iterating adversary and defender training in a adversarial loop, would yield defenders more robust to unseen strategies than detectors trained on static datasets.

Purely data-driven detectors are at a structural disadvantage against adversaries optimised to lie on the manifold of plausible trajectories. Physics-informed detectors, scoring deviations between observed joint dynamics and predictions, combined with cross-layer defences that jointly analyse telemetry and control signals, can additionally flag inconsistencies between the waypoints issued by the VE and the resulting joint behaviour, which is the discrepancy our attack introduces.

A full evaluation of these directions is beyond the scope of this paper, whose contribution is the attack itself. We note that the most robust deployments are likely to combine several approaches and we view our publicly released attack as a method on which this defensive research can be trained to protect against.


\subsection{Future Development}
Despite the promising results achieved in this study, several areas warrant further exploration. Future work should investigate the use of a diverse anomaly detection system that combines rule-based and advanced machine learning techniques to enhance robustness against adversarial attacks. 

Examining the impact of the attack on non-targeted joints would provide deeper insights into potential collateral mechanical stress. 
Given the widespread adoption of DT technology in industrial domains, ensuring the security and resilience of these systems against sophisticated adversarial threats is of paramount importance. By demonstrating the capabilities of DRL-driven adversarial strategies, this research aims to raise awareness and support the development of robust defensive measures that can effectively counter such intelligent, stealthy attacks. Additionally, this work aims to illustrate the threat of DRL-assisted attacks against DT-enabled systems and emphasises the need to train existing defence mechanisms using real network traffic and attack-driven logs within the DT pipeline. This also lays the foundation for future work that leverages these insights to strengthen and adapt current defence strategies, helping reduce the threat posed by DRL-assisted adversaries. We include a discussion on ethics and dual-use in Appendix \ref{sec:ethics}.

\section{Conclusions}
\label{sec:Conclusions}
This work presents a novel and covert wear-out attack using Deep Reinforcement Learning (DRL) to damage Digital Twin (DT)-enabled infrastructures. The adversarial agent strategically introduces perturbations to the control signals received by the physical robot, causing 54\% more mechanical stress in the targeted joint while evading detection by an ensemble of \acf{LSTM}, \acf{CNN} and \acf{ResNet} Autoencoder anomaly detectors. The agent's "low \& slow" approach allows it to maintain a higher torque at the target joint for long periods, thereby stealthily accelerating its degradation and increasing maintenance costs. The adversarial agent shows great effectiveness in a grey-box setting with minimal information and rapid adaptability to unforeseen environments, indicating how dangerous it can be. We hope that the proposed adversarial agent can be used by researchers to facilitate the design and development of robust defence measures that can prevent such covert, intelligent attacks.

\bibliographystyle{IEEEtran}
\bibliography{sample-base}

\appendices

\section{Acknowledgements}
This work was supported in part by the Engineering and Physical
Sciences Research Council (EPSRC) under Award EP/V039156/1.

\section{Ethical Discussion}\label{sec:ethics}
This work introduces an AI-assisted methodology for generating stealthy wear-out attacks against digital twin–enabled systems. While the primary objective is to advance defence mechanisms, we acknowledge the inherent dual-use risks associated with this framework and its potential for misuse.

The methods described in this paper could be repurposed by adversaries to conduct wear-out attacks against digital twin–enabled systems, increasing operational costs for targeted systems. Specifically, AI-driven optimisation of adjustments to operational commands, while maintaining stealth, amplifies these risks.

In recognition of these risks, we disclose all code and the datasets used to train both the anomaly detection models and the reinforcement learning environment. This enables security teams to develop defensive mechanisms against the proposed attack, improving overall system security. Additionally, we release our models to support the development of more advanced agents, encouraging continued research into RL-assisted adversaries and their role in strengthening system security.

In conclusion, we acknowledge the dual-use nature of the proposed methodology but contend that disclosing the model and data supports the security community in identifying emerging threats to digital twin systems and fosters deeper discussion on defending against AI-assisted attacks.

\section{IDS Performance Results}\label{app:IDS}

In this section, we present the full performance results of the \ac{IDS} used within the MuJoCo simulation environment that the experiments were performed in. Table~\ref{tab:metrics-avg} reports the performance of each detector averaged across the three robotic arm models used in this study (UR5e, UR10e, and Franka Panda), and Table~\ref{tab:metrics-time} reports performance in the time-enabled environment, in which timestamps are appended to joint-state sequences to provide resilience against replay-style attacks.

\subsection{Basic Environment}\label{sec:ids-basic}

Table~\ref{tab:metrics-avg} presents the results of the \ac{ResNet} autoencoder, \ac{CNN} autoencoder, and standard \ac{LSTM} models each achieve perfect scores across all five reported metrics (accuracy, F1, precision, recall, AUC), correctly classifying every sequence in the held-out evaluation set. Two interpretations are consistent with these results. The first is that the deviation between normal and adversarially modified trajectories is sufficiently large relative to the $\pm 0.1$\,mm Gaussian process noise applied during normal operation that the differences between decision boundaries is negligible for the presented architectures. The second is that the detectors have over-fit to the specific structure of the training trajectories; this would be expected to manifest as degraded performance under distribution shifts, and we observe partial evidence of this in the time-enabled environment. The two explanations are not mutually exclusive: \ac{ResNet} and \ac{CNN} architectures tend to fit narrow distributions tightly, and the fact that all three top-performing models converge on identical perfect scores is consistent with learning closely related decision boundaries on the same problem.

\begin{figure}[t!]
    \centering
    \begin{subfigure}{\linewidth}
        \centering
        \includegraphics[width=\linewidth]{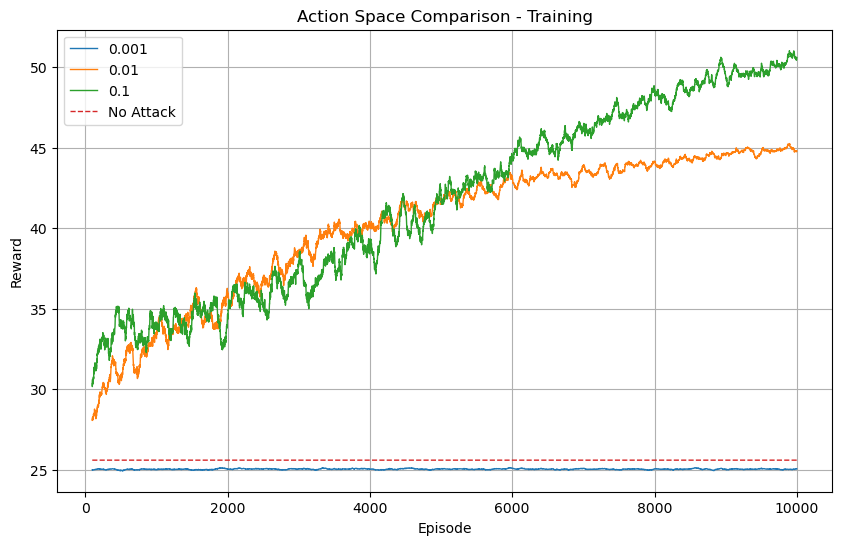}
        \caption{Training Rewards}
        \label{fig:action-space-training}
    \end{subfigure}
    \vspace{0.5em}
    \begin{subfigure}{\linewidth}
        \centering
        \includegraphics[width=\linewidth]{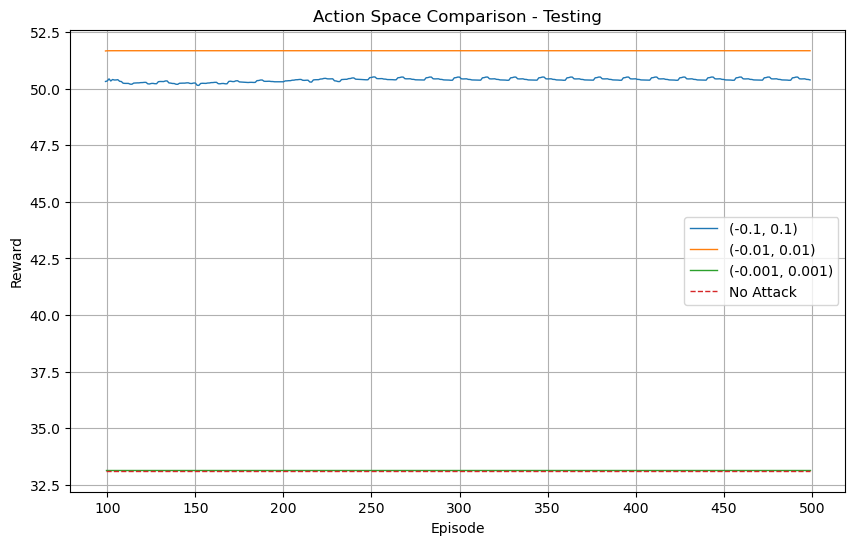}
        \caption{Testing Rewards}
        \label{fig:action-space-testing}
    \end{subfigure}
    \caption{Action Space Experimentation.}
    \label{fig:action-space}
\end{figure}

The \ac{GRU} and \ac{LSTM} autoencoders trail the top three. The \ac{GRU} autoencoder achieves $96.97\%$ accuracy, F1 of $0.9841$, precision of $0.9697$, recall of $1.0000$, and AUC of $0.9267$. The difference between precision and recall (a recall of $1.0000$ with precision below $0.97$) indicates that the model captures malicious sequences in the evaluation set but produces a small number of false positives on normal trajectories. For an \ac{IDS} deployed in an industrial setting where false alarms incur operational cost, this is the more tolerable failure mode of the two, missed detections would enable the attacks studied in this paper, whereas false alarms can be confirmed by operators. The corresponding AUC of $0.9267$ indicates that the \ac{GRU} autoencoder's reconstruction-error distribution still separates the two classes well across thresholds, but with less margin than the top three models.

The \ac{LSTM} autoencoder is the weakest detector in the basic environment, reporting $93.33\%$ accuracy, F1 of $0.9649$, precision of $0.9394$, recall of $0.9933$, and AUC of $0.7813$. The gap between recall ($0.9933$) and AUC ($0.7813$) is substantially larger than the other models which indicates that while the \ac{LSTM} autoencoder catches most adversarial sequences the underlying score distribution is poorly calibrated: small threshold shifts would produce large changes in the precision/recall trade-off. Second, the precision of $0.9394$ is the lowest of any model in either table, meaning the \ac{LSTM} autoencoder produces the highest false-positive rate at its chosen operating point. Together, these observations suggest the \ac{LSTM} autoencoder learns a less stable reconstruction of normal behaviour than other models.

\subsection{Time-Enabled Environment}\label{sec:ids-time}

Adding timestamps to the inputs produces small but consistent degradations in four of the five models (seen in Table~\ref{tab:metrics-time}). The \ac{CNN} autoencoder drops from $100\%$ to $98.55\%$ accuracy, the \ac{ResNet} autoencoder drops from $100\%$ to $99.88\%$ the \ac{GRU} autoencoder drops from $96.97\%$ to $96.36\%$ , and the \ac{LSTM} autoencoder drops from $93.33\%$ to $92.36\%$. The corresponding F1, precision, and AUC metrics shift in the same direction for each affected model, with AUC providing the most sensitive signal: the \ac{CNN} autoencoder's AUC falls from $1.0000$ to $0.9676$, and the \ac{ResNet} autoencoder's AUC falls from $1.0000$ to $0.9996$. The relative performance of the \ac{ResNet} autoencoder's AUC suggests the residual connections provide a stabilising effect on the additional input dimension, allowing the detector to incorporate timestamps without disrupting its learned representation of joint behaviours.

The standard \ac{LSTM} model is the only detector that maintains performance across both environments, reporting $100\%$ accuracy, F1, precision, recall, and AUC in both Table~\ref{tab:metrics-avg} and Table~\ref{tab:metrics-time}. This is a notable result: while the autoencoders treat timestamps as an additional reconstruction the \ac{LSTM} uses timestamps directly as a classification feature, gaining grounding without reducing reconstruction cost.

\begin{table}[t!]
\centering
\caption{Performance metrics (average over robotic arm models).}
\label{tab:metrics-avg}
\begin{tabularx}{\linewidth}{@{}l *{6}{Y}@{}}
\toprule
\multirow{2}{*}{\textbf{Metric}} & \multicolumn{5}{c}{\textbf{Model (Avg)}} \\ 
\cmidrule(lr){2-6}
 & \textbf{GRU AE} & \textbf{LSTM AE} & \textbf{ResNet AE} & \textbf{CNN AE} & \textbf{LSTM} \\ 
\midrule
Accuracy (\%)    & 96.97 & 93.33 & 100.00 & 100.00 & 100.00 \\
F1-Score         & 0.9841 & 0.9649 & 1.0000 & 1.0000 & 1.0000 \\
Precision        & 0.9697 & 0.9394 & 1.0000 & 1.0000 & 1.0000 \\
Recall           & 1.0000 & 0.9933 & 1.0000 & 1.0000 & 1.0000 \\
AUC              & 0.9267 & 0.7813 & 1.0000 & 1.0000 & 1.0000 \\
\bottomrule
\end{tabularx}
\end{table}

\begin{table}[t!]
\centering
\caption{Performance metrics (average over robotic arm models with time consideration).}
\label{tab:metrics-time}
\begin{tabularx}{\linewidth}{@{}l *{6}{Y}@{}}
\toprule
\multirow{2}{*}{\textbf{Metric}} & \multicolumn{5}{c}{\textbf{Model (Avg)}} \\ 
\cmidrule(lr){2-6}
& \textbf{GRU AE} & \textbf{LSTM AE} & \textbf{ResNet AE} & \textbf{CNN AE} & \textbf{LSTM} \\ 
\midrule
Accuracy (\%)  & 96.36 & 92.36 & 99.88 & 98.55 & 100.00 \\
F1-Score       & 0.9808 & 0.9594 & 0.9993 & 0.9921 & 1.0000 \\
Precision      & 0.9632 & 0.9366 & 0.9987 & 0.9845 & 1.0000 \\
Recall         & 1.0000 & 0.9853 & 1.0000 & 1.0000 & 1.0000 \\
AUC            & 0.9125 & 0.8174 & 0.9996 & 0.9676 & 1.0000 \\
\bottomrule
\end{tabularx}
\end{table}

\end{document}